\documentclass[aps,prb,reprint,groupedaddress]{revtex4-1}

\usepackage{graphicx}

\begin{document}

\title{ Domain State of the ANNNI model in Two Dimensions}

\author{Fumitaka Matsubara}%
\affiliation{%
Department of Applied Physics, Tohoku University, Sendai 980-8579, Japan%
}%
\author{Takayuki Shirakura}%
\email[]{shira@iwate-u.ac.jp}%
\affiliation{%
Faculty of Humanities and Social Sciences,%
Iwate University, Morioka 020-8550, Japan%
}%
\author{Nobuo Suzuki}%
\affiliation{%
Faculty of Science and Technology, Tohoku Bunka Gakuen University, Sendai 980-8551, Japan%
}%


\date{\today}

\begin{abstract}

We have examined the spin ordering of an axial next-nearest-neighbor 
Ising (ANNNI) model in two dimensions (2D) near above the 
antiphase ($\langle 2 \rangle$ phase). 
We considered an $N_R$-replica system and calculated an overlap function 
$q_m$ between different replicas having used a cluster 
heat bath (CHB) Monte Carlo method. 
We determined transition temperature between $\langle 2 \rangle$ phase 
and a floating incommensurate (IC) phase as $T_{C2}/J = 0.89 \pm 0.01$ 
with frustration ratio $\kappa (\equiv -J_2/J_1) = 0.6$. 
We found that the spin state at $T \gtrsim T_{C2}$ may be called 
as a domain state, 
because the spin structure is characterized by a 
sequentially arranged four types of domains with different 
$\langle 2 \rangle$ structures. 
In the domain state, the 2D XY symmetry of the spin 
correlation in the IC phase weakly breaks and the diversity of 
the spin arrangement increases as $T \rightarrow T_{C2}$. 
The Binder ratio $g_L$ exhibits a depression at $T \sim T_{C2}$ 
and the quasi-periodic spin structure, which is realized in the 
IC phase, becomes diverse at $T \gtrsim T_{C2}$. 
We discussed that the domain state is stable against the thermal 
fluctuation which brings a two-stage development of the spin structure 
at low temperatures.

\end{abstract}

\pacs{75.50.Lk,05.70.Jk,75.40.Mg}

\maketitle


\section{Introduction}


Systems with competitive interactions have been extensively studied throughout
 the past three decades, because they exhibit rich physical phenomena, 
 such as commensurate-incommensurate phase transitions, 
Lifshitz points and multiphase points.\cite {Selke0} 
The axial next-nearest-neighbor Ising (ANNNI) model is among the 
simplest realizations of such systems. 
In the two-dimensional (2D) ANNNI model, ferromagnetic Ising chains are 
coupled by ferromagnetic nearest-neighbor and antiferromagnetic 
next-nearest-neighbor interchain interactions on a square lattice. 
The Hamiltonian is described by
\begin{eqnarray}
\mathcal{H} &=& - \it{J}\sum_{\it{x,y}}S_{\it{x,y}}S_{\it{x+1,y}} \nonumber \\
     &-& \it{J_1}\sum_{\it{x,y}}S_{x,y}S_{x,y+1}
     - \it{J}_{\it{2}}\sum_{{x,y}}S_{x,y}S_{x,y+2},
\end{eqnarray}
where $S_{x,y}=\pm 1$ is an Ising spin. In this paper we consider the case 
with $J_1\; =\; J\; >0$ and $J_2 < 0$.
The ground state of the model is a ferromagnetic long range order (LRO) 
phase for frustration coefficient $\kappa (\equiv -J_2/J) <1/2$ 
and a LRO antiphase ($\langle 2 \rangle$ phase)  for $\kappa > 1/2$, 
in which the $\langle 2 \rangle$ phase is described by 
an alternate arrangement 
of two up-spin and two down-spin chains in the $y$-direction. 
This model at finite temperatures has been studied throughout the past few 
decades by various methods.\cite{Selke1,Selke2,Villain,Grynberg1,Saqi,Morita} 
It was suggested that, for $\kappa > 1/2$, a floating incommensurate (IC) 
phase exists between the $\langle 2 \rangle$ phase 
and the paramagnetic (PM) phase.\cite{Selke1,Selke2}
The IC phase close to the higher transition temperature, 
$T_{C1}$, may be characterized by dislocations\cite{Selke2} that play 
the same role of vortices in two-dimensional XY (2D XY) model,\cite{KT} 
and the IC phase near above the lower transition temperature,
 $T_{C2}$, may be characterized by domain walls of three up-spin chains 
or three down-spin chains penetrating the $\langle 2 \rangle$ phase. 
However, the spin ordering is yet to be clarified, 
because recent Monte Carlo (MC) studies predicted 
different spin orderings. 
While equilibrium MC simulations supported the above picture of 
the spin ordering with $T_{C2} < T_{C1}$,\cite{Sato,Rastelli}
a nonequilibrium relaxation (NER) MC method\cite{ItoA,ItoB} 
predicted $T_{C2} \sim T_{C1}$, i.e., the absence 
of the IC phase.\cite{Shirahata,Chandra}  
In a previous paper\cite{Shirakura} (referred as I, hereafter), 
we reexamined the spin ordering of the ANNNI model with $\kappa = 0.6$ 
near $T_{C1}$ having used both the equilibrium MC 
and the NER MC methods and showed that both methods give almost 
the same transition temperature of $T_{C1} \sim 1.16J$ 
and the spin ordering at $T \lesssim T_{C1}$ exhibits properties 
of the 2D XY model.\cite{KT} 

In this paper, we reexamine the spin ordering near 
the other transition temperature $T_{C2}$. 
We will propose a physical quantity appropriate for the ANNNI model, 
by which we can readily separate $\langle 2 \rangle$ phase from 
the IC phase. 
Section	II describes the investigated method and quantities 
of the 2D ANNNI model. 
Section III presents results of the equilibrium simulation. 
In Sec.~IV, we discuss, on the basis of a domain picture, 
the spin structure of the model for both the temperature ranges 
of near below $T_{C1} (T \lesssim T_{C1})$ 
and near above $T_{C2} (T \gtrsim T_{C2})$.  
Section V is addressed on the periodic nature of the ANNNI model 
calculating the Fourier component of the spin arrangement. 
Section VI is devoted to conclusions and discussions, 
where we will discuss why the 2D ANNNI model undergoes 
a very slow relaxation at low temperatures.


\section{Method and Quantities}


We apply a similar technique as proposed by 
Sato and Matsubara(SM)\cite{Sato} and used in I. 
We consider the model with $\kappa = 0.6$ on the $L_0 \times L_0$ lattice 
with open boundaries. 
Quantities of interest are measured in the inner region of $L \times L$ 
with $L = L_0/2$ , i.e., $L_0/4+1 \leq x, y \leq 3L_0/4$. 
Hereafter we attach a new lattice site name for this region, i.e., 
$x,y = 1, 2, \cdots, L$. 
We apply the CHB algorithm\cite{CHB1,CHB2} to get an equilibrium 
spin configuration. 
That is, the spin configuration of a block of $L_0 \times l_y$ spins 
is updated using the transfer matrix method, 
where the transfer direction is the $x$-direction ($L_0$) and the width 
of the block $l_y$ is determined from the computational time costs. 
In the previous paper we apply the SM procedure with $l_y=6$. 
The choice of $l_y=6$ was appropriate in a temperature range of 
$T \sim T_{C1}$. However, when the temperature is lowered toward $T_{C2}$, 
the number of the MC sweeps needed to get the equilibrium spin configuration 
rapidly increases. 
Then we adopt a larger $l_y$ for a larger lattice of $L_0 \times L_0$.

The difficulty of studying the phase transition of the ANNNI model 
is that the spin structure of the IC phase is not known a priori. 
Another problem to be noted is that the $\langle 2 \rangle$ phase has 
equivalent four structures. 
That is, when we look at four spins at $(4l+1, 4l+2,4l+3,4l+4)$ 
($l = 0,1,2 \cdots$) chains, they have either 
$(++--)$, $(+--+)$, $(--++)$ or $(-++-)$. 
One usually investigates the squared chain magnetization $M_2 
(= \frac{1}{L}\sum _{y=1}^{L} (\frac{1}{L} \sum _{x=1}^{L} S_{x,y})^2)$. 
This quantity characterizes the spin correlation along the $x$-direction. 
Using $M_2$, one can separate the IC phase from the PM phase, 
because in the PM phase $M_2$ exponentially decays with increasing $L$  
and $M_2$ will algebraically decays in the IC phase. 
However, near the lower transition temperature $T_{C2}$, calculation 
of equilibrium value of $M_2$ for a larger size $L_0$ is hard 
task\cite{Sato,Shirakura} and one searches for $T_{C2}$ 
using different quantities.\cite{Sato,Shirahata}
Moreover we can hardly investigate the 2D natue of the ANNNI model, 
in particular the spin correlation along the $y$-direction.

Here we consider another quantity for examining the 2D spin structure itself. 
We obtain the equilibrium spin configuration $\{ S^0_{x,y} \}$ 
at a temperature $T$. 
That is, we get $\{ S^0_{x,y} \}$ performing the MC simulation 
using some sequence of random numbers. 
Then, make the MC simulation at the same temperature $T$ using 
a different sequence of random numbers and get the spin 
configuration $\{ S_{x,y} \}$. We extract the $\{ S^0_{x,y} \}$ component 
of $\{ S_{x,y} \}$ by calculating the spin overlap between them. 
To realize this procedure, we consider an $N_R$-replica system. 
The spin configurations $\{S_{x,y}^{\alpha}\}$ ($\alpha = 1, 2, \cdots, N_R$) 
of these replicas are generated by different sequences of random numbers. 
We define the $\vec{k}$-dependent maximum spin overlap function 
$q_m^{\alpha,\beta}(\vec{k})$ of the replica $\alpha$ and 
the replica $\beta$ as 
\begin{eqnarray}
q_m^{\alpha,\beta}(\vec{k})= q^{\alpha,\beta}(y_0,\vec{k}), 
\end{eqnarray}
where 
\begin{eqnarray}
y_0 = \arg\max_{-L/2 \leq y' \leq L/2} q^{\alpha,\beta}(y',\vec{k}), 
\hspace{0.8cm}  \\
q^{\alpha,\beta}(y',\vec{k})=\frac{1}{L^2} |\sum_{x=1}^{L}\sum_{y=1}^{L}
S_{x,y}^{\alpha}S_{x,y+y_0}^{\beta}\exp(i\vec{k}\vec{r}_{x,y+y'})|.
\hspace{0.2cm} 
\end{eqnarray}
In the IC phase a drift of the spin configuration inevitably occurs 
along the $y$-direction. We take into account the drift 
by a uniform shift of the spin configuration $y_0$. 
Also $q_m^{\alpha,\beta}(\vec{k})$ is free from the structure of the 
$\langle 2 \rangle$ phase.
The overlap function $q_m(\vec{k})$ of the system is the average 
of those overlap functions.  
\begin{eqnarray}
q_m(\vec{k}) = \frac{2}{N_R(N_R-1)} \sum_{\alpha \neq \beta}
q_m^{\alpha,\beta}(\vec{k})
\end{eqnarray}
Note that the overlap function at $\vec{k} = 0$, $q_m (\equiv q_m(0))$, 
plays the role of the order parameter of the ANNNI model. 
If $q_m$ decys algebraicly with increasing $N$, it reveals that the system 
is in the critical phase, i.e., the IC phase, while $q_m$ remains non-zero 
constant, it reveals that the system is in the LRO phase, 
i.e., the $\langle 2 \rangle$ phase. 
In contrast, $q_m(\vec{k})$ for $\vec{k} \neq 0$ will be used to 
obtain the correlation length of the spin structure. 


\section{Results} 


We investigate the equilibrium properties of the ANNNI model by the 
overlap function $q_m$. 
We focus our effort in the temperature range of $T \leq T_{C1} (\sim 1.16J)$.
We perform the CHB simulation of the ANNNI model on lattices with 
$L_0 = 24 \sim 128$. 
We make two simulations: a gradual cooling simulation 
and a gradual heating simulation. 
In the gradual cooling (heating) simulation, we start with a 
PM ($\langle 2 \rangle$) spin configuration at a high (low) 
temperature and perform the simulation described below,
then the temperature is lowered (raised) by some fixed interval 
$\Delta T$ and perform the same simulation starting with the last 
spin configuration at the previous temperature, and so on. 
For each temperature, after $MCS_{\rm equi}$ sweeps are discarded, 
data of interest are measured for every 10 sweeps over 
$MCS_{\rm mea}$ sweeps. 
Data presented here are averages of those of two simulations, 
and errors are differences between them. 
Hereafter averages of data $Q$ in the inner $L \times L$ lattice 
are described as $\langle Q \rangle_L$. 
The parameters used in the equilibrium simulation are listed in 
Table~\ref{tab:table1}.

\begin{table}[b]
\caption{\label{tab:table1}
Parameters used in the CHB algorithm of the MC simulation of 
the $N_R = 8$ replica system. 
$MCS_{\rm equi}$ and $MCS_{\rm mea}$ are the number of MC sweeps required 
for equilibration and measurement, respectively.}
\begin{ruledtabular}
\begin{tabular}{rrrr}
$L_0$ &$l_y$ & $MCS_{\rm equi}$ & $MCS_{\rm mea}$ \\ 
\hline
  $\leq$ 48 & 6 & 4,000 & 12,000 \\ 
  64 &8 & 10,000 & 30,000 \\
 96 &10 &  20,000 & 60,000 \\ 
 128 &12 &  40,000 & 80,000 \\
\end{tabular}
\end{ruledtabular}
\end{table}


\subsection{Spin overlap }


Figure 1 shows $\langle q_m \rangle_L$ as functions of $T$ for different $L$. 
At high temperatures, $\langle q_m \rangle_L$ for a larger $L$ is 
smaller than that for a smaller $L$.
As the temperature is decreased from a high temperature, 
$\langle q_m \rangle_L$'s for all $L$ increase and come together 
at $T \sim 0.89J$. Below this temperature, the $L$-dependence of 
$\langle q_m \rangle_L$ is reversed. 
This result clearly reveals that some 2D LRO takes place at $T < 0.89J$. 
That is, the transition temperature between the IC phase 
and the $\langle 2 \rangle$ phase is $T_{C2}/J = 0.89 \pm 0.01$, 
because the LRO phase of the 2D ANNNI model is the $\langle 2 \rangle$ phase.

\begin{figure}
\includegraphics{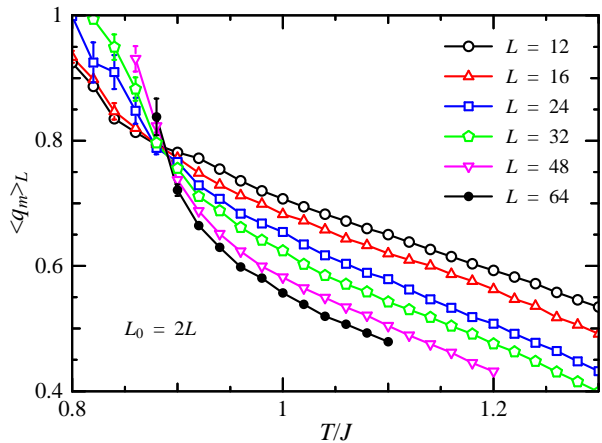}%
\vspace{-0.4cm}
\caption{ (color online) 
Temperature dependences of the spin overlap function $\langle q_m \rangle_L$ 
in the ANNNI model with $\kappa = 0.6$, computed over the inner lattice with 
$L = L_0/2$. }
\end{figure}


\subsection{Binder Ratio}


Next we consider the Binder ratio\cite{Binder} 
to examine the nature of the phase transition at $T_{C2}$. 
In a usual ferromagnetic(FM) model on a cubic lattice 
with the linear dimension $L$, 
the Binder ratio $g_L$ of the magnetization $M$ monotonically 
increases with decreasing temperature 
and reach 1 for $T \rightarrow 0$. 
In the PM phase, $g_L$ decreases with increasing $L$, 
while $g_L$ increases with $L$ in the FM phase. 
Then the Binder ratio is independent of $L$ 
at the critical temperature $T = T_C$. 
That is, the Binder ratios of different $L$'s intersect at $T = T_C$. 

The Binder ratio $g_L$ of $q_m$ is defined as
\begin{eqnarray}
g_L = \frac{1}{2}\left(3-\frac{\langle q_m^4\rangle_L}{\langle q_m^2\rangle_L^2}\right).  
\end{eqnarray}
Figure 2 shows temperature dependences of $g_L$ for different $L$. 
They show unusual behaviors. 
For $T > T_{C2}$, $g_L$ increases with $L$ and seems to reach some finite 
value which is smaller than 1. 
On the other hand, for $T < T_{C2}$ $g_L$ rapidly increases toward 1. 
This result is compatible with the fact that the IC phase for 
$T_{C2} < T < T_{C1}$ is the KT-like phase, i.e., a critical state, 
and that the $\langle 2 \rangle$ phase for $T < T_{C2}$ is the LRO phase. 
A queer point is its temperature dependence.  
As the temperature is decreased from a high temperature, 
$g_L$ once increases, reaches its maximum value at 
$T \sim 1.00J$, then decreases down to around $T_{C2}$. 
That is, the phase transition at $T_{C2}$ accompanies with 
a diversity of the spin structure. 

\begin{figure}
\includegraphics{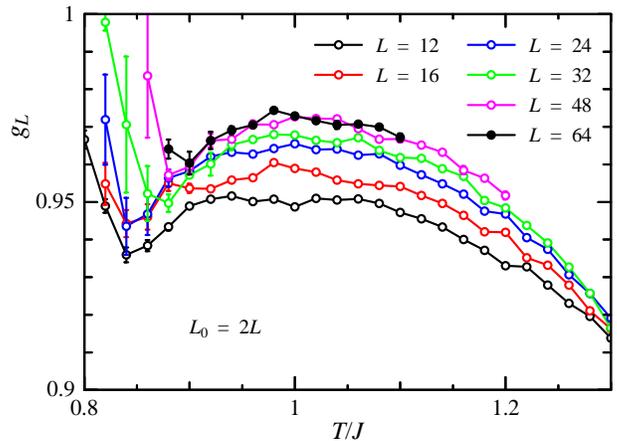}\\
\caption{(color online) 
Temperature dependences of Binder ratio $g_L$ in the ANNNI model 
with $\kappa = 0.6$ and different $L$.}
\end{figure}

\begin{figure}
\includegraphics[width=7.0cm]{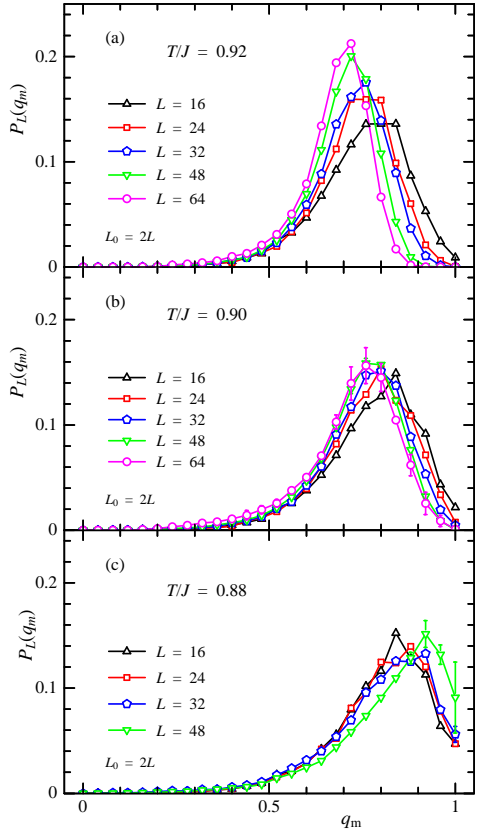}\\
\vspace{-0.4cm}
\caption{(color online) 
Distribution $P_L(q_m)$ of the spin overlap $q_m$ 
in the ANNNI model with $\kappa = 0.6$. }
\end{figure}

We consider the distribution $P_L(q_m)$ of the order parameter $q_m$ 
to investigate the diversity of the spin structure. 
Attention is paid whether $P_L(q_m)$ exhibits a usual single peak 
reminiscent of the continuous phase transition 
or a double peak of the first order phase transition. 
Figures 3(a)-(c) show $P_L(q_m)$ for $T \sim T_{C2}$. 
They exhibit a single peak revealing that 
the phase transition at $T_{C2}$ is some continuous one. 
For $T > T_{C2}$, as $L$ increases, the peak position $q_m^{(p)}$ 
shifts to the small $q_m$ side and the peak height $P_L(q_m^{(p)})$ 
seems to saturate. 
These results are compatible with the $L$-dependences of 
$\langle q_m \rangle_L$ and $g_L$ shown in Figs. 1 and 2, respectively. 
As $T \rightarrow T_{C2}$, $P_L(q_m)$ becomes broader and 
seems to be independent of $L$. 
For $T < T_{C2}$, the weight of $P_L(q_m)$ at smaller $q_m$ diminishes and 
$P_L(1)$, which is the weight of the $\langle 2 \rangle$ phase, increases. 
Therefore the depression of $g_L$ at $T \sim T_{C2}$ corresponds with a 
spread of $P_L(q_m)$. 
We believe the spread attributes to a characteristic nature intrinsic 
to the 2D ANNNI model. 
We consider the $\langle 2 \rangle$ phase with some $\langle 2 \rangle$ 
structures, i.e., a single-domain state $\{S_{x,y}^{\alpha}\}$. 
Suppose that one domain wall penetrates in the system. 
Then the system is separated into two domains with different 
$\langle 2 \rangle$ structures, i.e., 
a two-domain state $\{S_{x,y}^{\beta}\}$. 
The spin overlap function $q_m^{\alpha,\beta}(\vec{k})$ 
takes various values, depending on the location and the shape 
of the domain wall. 
Therefore the broad peak of $P_L(q_m)$ at $T \gtrsim T_{C2}$ 
may suggest that the system is composed of domains 
with different $\langle 2 \rangle$ structures. 
In this temperature range, the spin correlations along 
the $x$- and $y$-directions will become anisotropic.

\subsection{Correlation length}

\begin{figure}
\includegraphics[width=8cm]{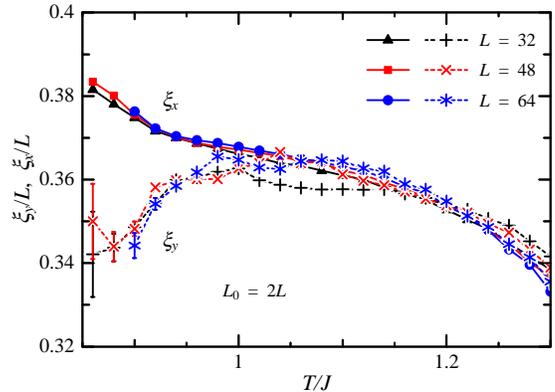}\\
\caption{(color online) 
Temperature dependences of the correlation-length ratios 
$\xi_x/L$ (a) and $\xi_y/L$ (b) in the $x$ and $y$ directions,
respectively, in the ANNNI model with $\kappa = 0.6$. }
\end{figure}

We consider the spin correlation length $\xi_{\mu}$ along the 
$\mu$ direction ($\mu = x, y$) to examine the speculation mentioned above. 
This quantity is obtained from the spin overlap function as follows:  
\begin{eqnarray}
\xi_{\mu} = \frac{1}{2\sin(|\vec k_{\rm min}|/2)}
\sqrt{\frac{\langle q_m^2\rangle_L}
{\langle |q({\vec k}_{\rm min})|^2\rangle_L}-1}
\end{eqnarray}
where $ {\vec k}_{\rm min} = (\pi/L, 0)$ and $ {\vec k}_{\rm min} = (0, \pi/L)$ 
in the $x$ and $y$ direction, respectively. 
One usually studies the ratio of the correlation length $\xi_{\mu}$ to 
the linear lattice size $L$,  $\xi_{\mu}/L$, to determine the transition 
temperature $T_C$.\cite{Cooper} 
Here we pay attention to the relation in the spin correlation 
between the $x$  and $y$ directions. 

Figure 4 shows the correlation-length ratios $\xi_x/L$ 
and $\xi_y/L$ for different $L$ as functions of $T$. 
In the $x$ direction $\xi_x/L$ smoothly increases with decreasing 
 temperature. 
On the other hand, in the $y$ direction $\xi_y/L$ exhibits 
an interesting behavior. 
As the temperature is decreased from a high temperature, 
$\xi_y/L$ once increases, becomes maximum at $T \sim 1.00J$,
then decreases down to $T_{C2}$. 
We note that the temperature dependence of $\xi_y/L$ is quite similar 
to that of $g_L$ shown in Fig. 2. 
This fact indicates that the diversity of the spin structure suggested 
by $g_L$ and $P_L(q_m)$ comes from the decrease of the spin correlation 
along the $y$ direction.

The remarkable point is that the nature of the spin correlation changes 
as the temperature is decreased toward $T_{C2}$. 
Of course, $\xi_y \sim \xi_x$ in the PM phase ($T > T_{C1} \sim 1.16J$). 
This relation holds near below $T_{C1}$ and remains down to $T \sim 1.0J$. 
That is, for $T > 1.0J$ the system is almost isotropic for every direction 
like that of the KT phase in the 2D XY model.
For $T \lesssim 1.0J$, $\xi_x$ and $\xi_y$ exhibit different 
temperature dependences, i.e., one increases with decreasing 
 temperature and the other decreases. 
Therefore $T = 1.0J$ is a temperature below which the nature of 
the spin correlation gradually changes from that of the KT-like 
state to that of another state, probably a domain state.


\section{Domain Structure}


\begin{table}[b] 
\caption{\label{tab:table2} 
Spin arrangements of the domain element {\bf \{$S_{x,y}$\}} $\equiv$
$(s_{x,y}, s_{x,y+1}, s_{x,y+2},s_{x,y+3})$ and the domain values 
 $\tau_{x,y} ( = a, b, \bar{a}$ or $\bar{b})$. Here the location of 
the element is distinguished by $y_0 = {\rm mod}[y,4]$.}
\begin{ruledtabular}
\begin{tabular}{cccc|cccc}
\multicolumn{4}{c|}{{\bf \{$S_{x,y}$\}} \ \  $\setminus$ \ \ $y_0$ } & 1 & 2 &  3 & 0 \\
\hline
 $+1$ & $+1$ & $-1$ & $-1$ & $a$ & $\bar{b}$ & $\bar{a}$ & $b$ \\
\hline
 $+1$ & $-1$ & $-1$ & $+1$ & $b$ & $a$ & $\bar{b}$ & $\bar{a}$ \\
\hline
 $-1$ & $-1$ & $+1$ & $+1$ & $\bar{a}$ & $b$ & $a$ & $\bar{b}$ \\
\hline
 $-1$ & $+1$ & $+1$ & $-1$ & $\bar{b}$ & $\bar{a}$ & $b$ & $a$ \\
\end{tabular}
\end{ruledtabular}
\end{table}

Now we examine the spin structure itself at $T \gtrsim T_{C2}$. 
Here we discuss it on the basis of the domain picture. 
For this aim, we define a domain valuable $\tau_{x,y} 
(=a, b, \bar{a}$ or $\bar{b})$ which describes the element of the domain. 
Values of $\tau_{x,y}$ are determined as follows. 
We consider sequential four spins ($s_{x,y}, s_{x,y+1}, s_{x,y+2},s_{x,y+3}$). 
Elements of the $\langle 2 \rangle$ structure are $(+1,+1,-1,-1)$, 
$(+1,-1,-1,+1)$, $(-1,-1,+1,+1)$, and $(-1,+1,+1,-1)$. These elements 
are distinguished by the location in the lattice. 
Since the $\langle 2 \rangle$ phase has the translational symmetry of 
$y \rightarrow y+4l$ $(l=1,2,\cdots)$, if the spin arrangement of 
$(s_{x,y+4l}, s_{x,y+4l+1}, s_{x,y+4l+2},s_{x,y+4l+3})$ is 
the same as that of $(s_{x,y}, s_{x,y+1}, s_{x,y+2},s_{x,y+3})$, 
$\tau_{x,y+4l} = \tau_{x,y}$. 
The spin configurations and domain values 
are listed in Table~\ref{tab:table2}. 
Hereafter we describe the domain composed of $d (=a, b, \bar{a}$ or $\bar{b})$ 
element as $D (= A, B, \bar{A}$ or $\bar{B})$ domain, respectively, 
and the domain wall between the $D_1$ and $D_2$ domains as $W_{D_1-D_2}$, 
where $D_1,D_2 = A, B, \bar{A}, \bar{B}$. 
When the D domain covers the whole lattice, we call the state  
the D-type $\langle 2 \rangle$ phase. 
We readily find an interesting property of the arrangement of neighboring 
two domains. 
If the domain wall is composed of three up-spin or down-spin chains, 
when we watch the domain with increasing the chain site $y$, 
the $A$ follows the $B$, the $B$ the $\bar{A}$, 
the $\bar{A}$ the $\bar{B}$, and the $\bar{B}$ the $A$. 
That is, the domains will appear sequentially as 
$A \rightarrow \bar{B} \rightarrow \bar{A} \rightarrow B 
\rightarrow A \rightarrow \bar{B} \cdots$.
An example of this is as follows.
\[\begin{array}{c|cccccccccccc}
y         & 1 & 2 & 3 & 4 & 5 & 6 & 7 & 8 & 9 & 10& 11& 12 \\
\hline
y_0       & 1 & 2 & 3 & 0 & 1 & 2 & 3 & 0 & 1 & 2 & 3 & 0 \\
\hline
S_{x,y}   & +1&+1&-1&-1&+1&+1&+1&-1&-1&+1&+1&-1 \\
\hline
\tau_{x,y} &  a&  a&  a& -& -& \bar{b}& \bar{b}& \bar{b}& \bar{b}& 
\bar{b}& - &- \\
\end{array}\]
\vspace{-0.6cm}
\[\begin{array}{cccccccccccccc}
      13 & 14 & 15 & 16 & 17 & 18 & 19 & 20 & 21 & 22 & 23 & 24 & 25 \\
\hline
      1 & 2 & 3 & 0 & 1 & 2 & 3 & 0 & 1 & 2 & 3 & 0 & 1 \\
\hline
      -1&-1&+1&+1&-1&-1&+1&+1&+1&-1&-1&+1&+1 \\
\hline
      \bar{a}& \bar{a}& \bar{a}& \bar{a}& \bar{a}& -& -& b& b& b& \cdot &
 \cdot & \cdot \\
\end{array}\]

Figures 5(a) and (b) shows the snapshots of the domain structures 
at $T \lesssim T_{C1}$ and at $T \gtrsim T_{C2}$, respectively.
For $T \lesssim T_{C1}$, the system is composed of small domains 
which are separated by tangled domain walls. 
On the other hand, for $T \gtrsim T_{C2}$, the system is composed of 
several large domains each of which runs across the lattice. 
That is, the difference in the domain structure between 
the two temperature ranges are the size of the domains. 
For either case, four types of the domains appear in order 
as speculated above.
We calculate the average domain width $W_L$ for different sizes $L$ 
of the lattice 
and extrapolate it to $L \rightarrow \infty$. 
Figure 6 shows $W_L$ as functions of $T$ together with its extrapolation. 
As the temperature is decreased from a high temperature, 
$W_L$ first increases slowly down to $T \sim 1.00J$, and then $W_L$ 
increases rapidly and its extrapolation seems to diverge as 
$T \rightarrow T_{C2}$.

\begin{figure}
\includegraphics[width=6cm]{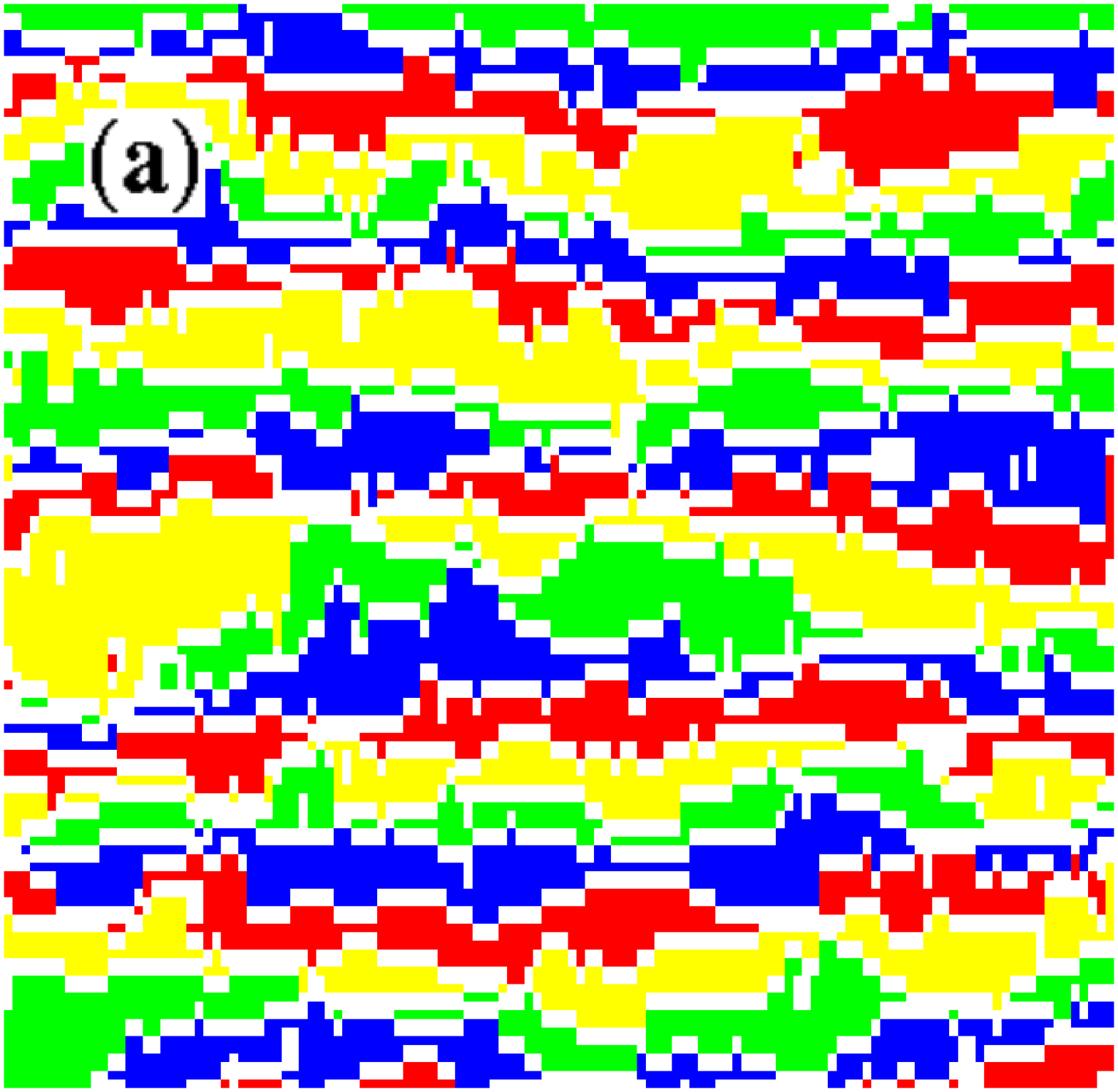}\\
\vspace{5mm}
\includegraphics[width=6cm]{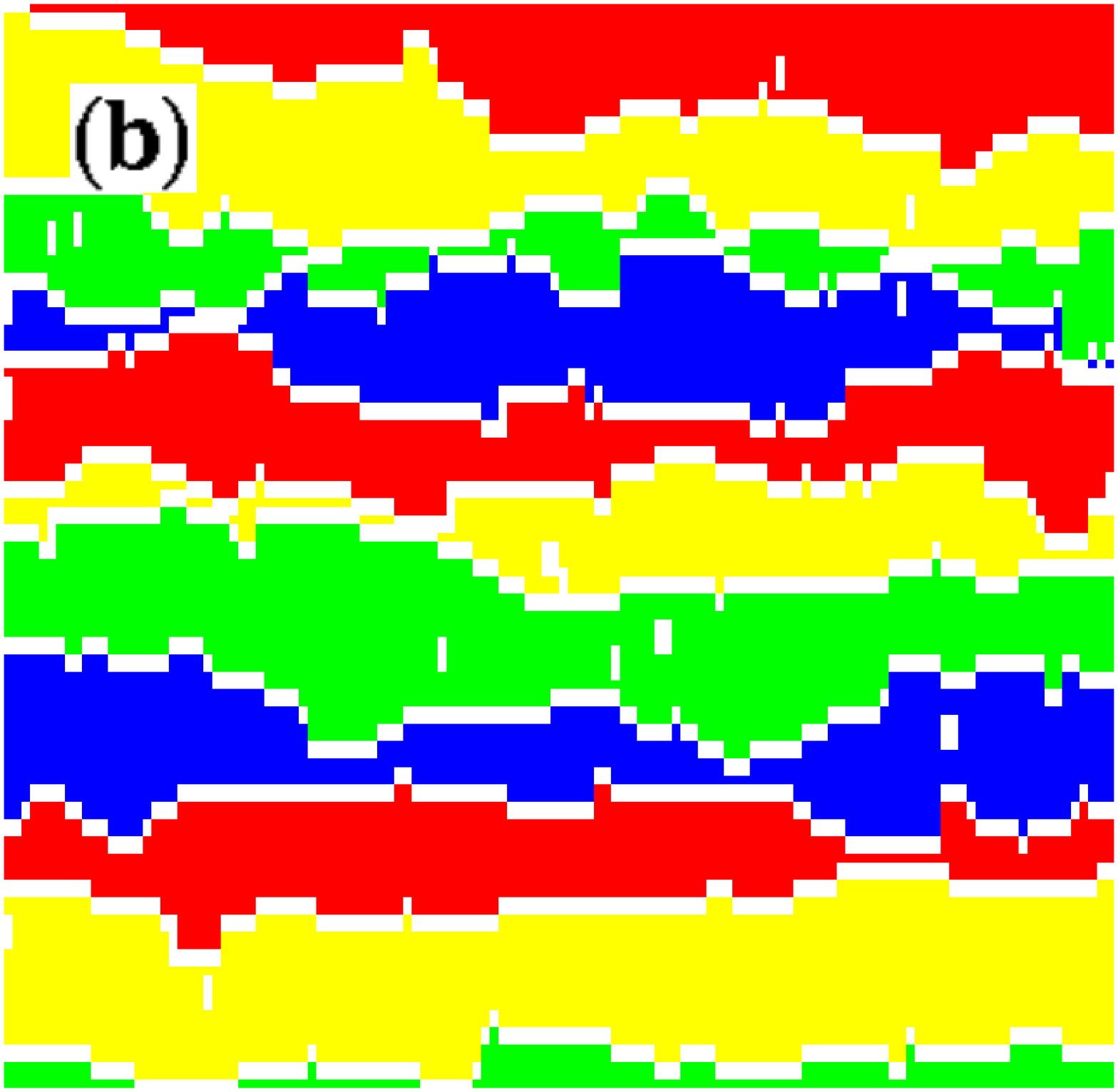}
\caption{(color online) 
Snapshots of the domain structure at (a) $T = 1.1J$ and (b) $T = 0.92J$ 
in the ANNNI model with $\kappa = 0.6$ on the $128 \times 128$ lattice.
The domain elements $a$, $b$, $\bar{a}$, and $\bar{b}$ are described 
by red, blue, green, and yellow, respectively, and the domain wall element 
by white. 
}
\end{figure}

\begin{figure}
\includegraphics{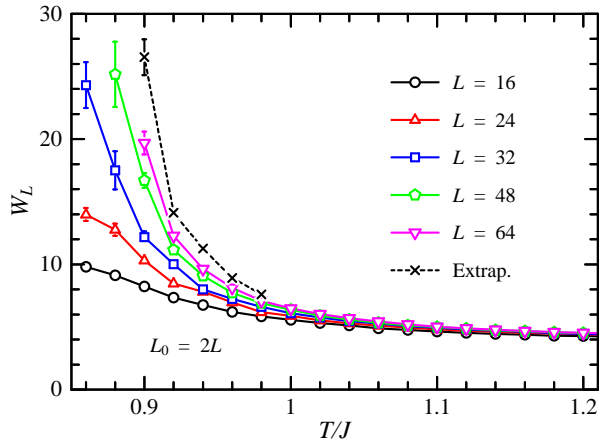} 
\caption{(color online) 
Temperature dependences of the average domain width $W_L$ 
of the lattice with $L$ 
in the ANNNI model with $\kappa = 0.6$. 
The extrapolation to $L \rightarrow \infty$ is made using 
$W_L = W_{\infty} + A/L$ assumption. 
}
\end{figure}


\section{Fourier Component $S(k)$}


\begin{figure}
\includegraphics{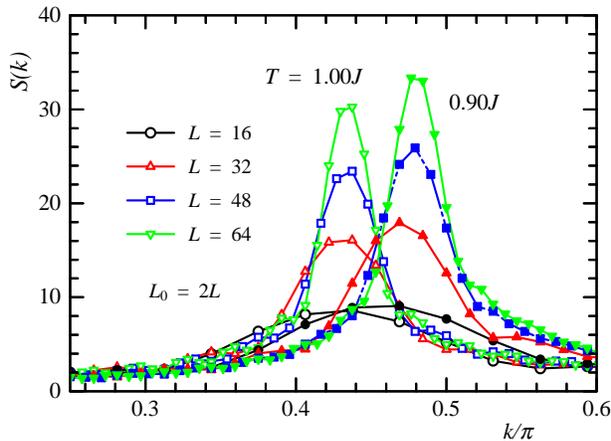} 
\caption{(color online) 
The Fourier components of the spin structure $S(k)$ for two temperatures 
well-above (open symbols) and near-above (solid symbols) $T_{C2}$. }
\end{figure}

\begin{figure}
\includegraphics{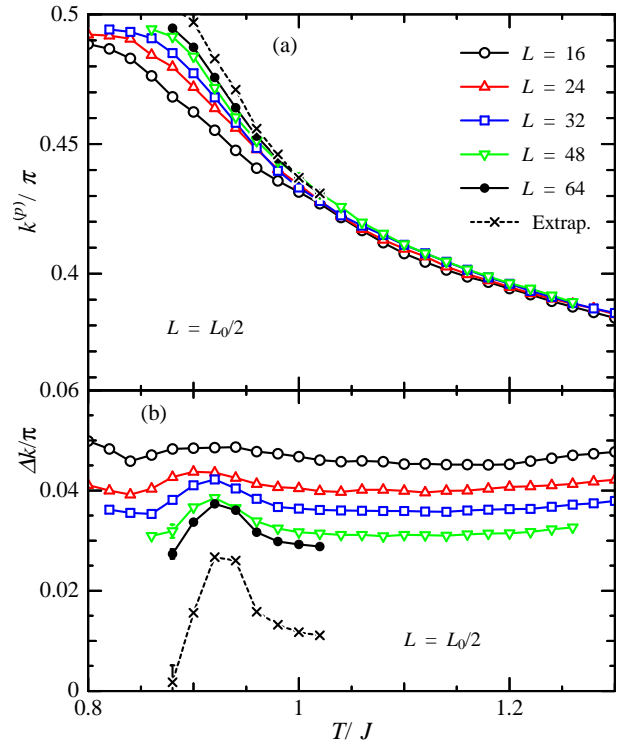} 
\caption{(color online) 
Temperature dependences of $k^{(p)}$ and $\Delta k$ 
of the lattice with $L$ in the ANNNI model with $\kappa = 0.6$. 
The extrapolations to $L \rightarrow \infty$ for $k^{(p)}$ and $\Delta k$ 
are made using $k^{(p)}_L = k^{(p)}_{\infty} + A/L$ and $\Delta k_L
= \Delta k_{\infty} + B/\sqrt{L}$ assumptions, respectively. 
}
\end{figure}

In an experimental point of view, the Fourier component of the spin 
arrangement is interesting. 
The Fourier component $S(k)$ along the $y$ direction is given by 
\begin{eqnarray}
 S(k) = |\sum _{y=1}^{L} m(y)\exp(iky)|,
\end{eqnarray}
where $m(y) (= \sum _{x=1}^LS_{x,y}/L)$ is the averaged magnetization 
of the $y$-th chain. 
This quantity gives the periodicity of the spin arrangement 
in the $y$ direction. 
The period $n$ of the spin arrangement is given by 
$n = 2\pi/k$ and the $\langle 2 \rangle$ phase corresponds to 
$k = \pi/2$. 
Figure 7 shows $S(k)$ at two temperatures of well-above 
and near-above $T_{C2}$ for different size $L$. 
For either temperature, $S(k)$ exhibits a single rather broad peak which 
grows with increasing $L$. 
The result suggests that the system has 
a quasi-periodic spin arrangement characterized by the peak position 
$k^{(p)}$ and its deviation $\Delta k$. 
Note that the number of the periodic spin arrangements describing $S(k)$ 
is roughly given by $n_p \sim \Delta k/(\pi/L) = (\Delta k/\pi)L$.  
Thus $\Delta k$ gives the diversity of the spin structure. 
Here we estimate the peak $k^{(p)}$ and its deviation $\Delta k$ from 
\begin{eqnarray}
      k^{(p)} &=& \sum_{k=\pi/L}^{\pi}k\tilde{S}(k)
                                /\sum_{k=\pi/L}^{\pi}\tilde{S}(k)\\
     (\Delta k)^2 &=& \sum_{k=\pi/L}^{\pi}(k-k^{(p)})^2\tilde{S}(k)
                                /\sum_{k=\pi/L}^{\pi}\tilde{S}(k).
\end{eqnarray}
where $\tilde{S}(k) = S(k)-\sum_{k=\pi/L}^{\pi}S(k)/{L}$.
We calculate $k^{(p)}$ and $\Delta k$ for different $L$ and extrapolate 
them to $L \rightarrow \infty$. 
Figures 8(a) and (b) show $k^{(p)}$ and $\Delta k$ for different $L$ as 
functions of $T$, respectively, together with their extrapolated values.
As the temperature is decreased from a high temperature, 
$k^{(p)}$ increases toward $k^{(p)} = \pi/2$ at $T_{C2}$. 
On the other hand, $\Delta k$ changes a little above $T_{C2}$ and 
the extrapolation value has a finite value, suggesting that 
some quasi-periodic spin structure occurs above $T_{C2}$. 
A notable thing is that $\Delta k$ has a mound near above $T_{C2}$. 
Again we see that the diversity of the spin structure along 
the $y$-direction is enhanced at $T \gtrsim T_{C2}$. 
For $T < T_{C2}$, $\Delta k$ diminihes revealing that 
the spin structure in the $\langle 2 \rangle$ phase 
is periodic with period four.

\section{Discussion and conclusions}

We have examined the spin ordering near the lower transition temperature 
$T_{C2}$ of the 2D ANNNI model with $\kappa = 0.6$ having used 
a CHB Monte Carlo method.
We considered an $N_R$-replica system and calculated an overlap function 
$q_m$ between different replicas. 
We determined $T_{C2}/J = 0.89 \pm 0.01$ and examined the nature of the 
spin structure at $T \gtrsim T_{C2}$ by the use of different quantities. 
The results were summarized in Fig. 9. 
In the floating IC phase, the nature of the spin correlation for 
$T \gtrsim T_{C2}$ are considerably different from that for 
$T \lesssim T_{C1}$. 
For $T \gtrsim T_{C2}$, the system is characterized 
by large domains each of which run across the lattice. 
Therefore we may call the spin structure for $T \gtrsim T_{C2}$  
a domain state. 

\begin{figure}
\includegraphics[width=7.0cm]{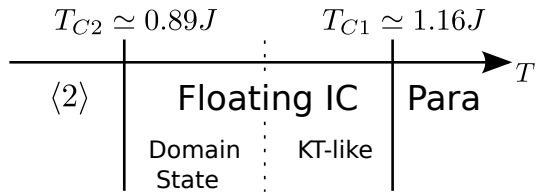}\\
\caption{(color online) 
The spin ordering of the 2D ANNNI model with $\kappa = 0.6$}
\end{figure}

In the domain state in Fig. 9, the 2D XY symmetry of the spin 
correlation breaks, i.e., the spin correlation length in the axial 
direction $\xi_y$ decreases with lowering temperature in contrast 
with a monotonous increase of that in the chain direction $\xi_x$. 
In consequence, the diversity of the spin arrangement increases 
as  $T \rightarrow T_{C2}$ and the Binder ratio $g_L$ exhibit a 
depression at $T \sim T_{C2}$. 
Also the quasi-periodic spin structure, which is realized in the 
IC phase, becomes diverse at $T \gtrsim T_{C2}$.

Here we consider properties of the domain state. 
We first note that an isolated domain 
is unstable, because it readily collapses with thermal noise. 
On the other hand, the domain state occurring at $T \gtrsim T_{C2}$ 
is sequentially arranged four types of domains: 
$A \rightarrow \bar{B} \rightarrow \bar{A} \rightarrow B 
\rightarrow A \rightarrow \bar{B} \cdots$. 
This structure is stable against the thermal fluctuation. 
We consider the $A$-type $\langle 2 \rangle$ phase with 
three domains of $\bar{B}, \bar{A}$, and $B$. 
Suppose that the $\bar{B}$ domain collapses. 
Then domain arrangement becomes as $A \rightarrow \bar{A} \rightarrow B 
\rightarrow A$ and an unfavorable domain wall of $W_{A-\bar{A}}$ appears. 
This domain structure is unstable and as soon as the $\bar{B}$ 
collapses, another $\bar{B}$ arises between the $A$ and the $\bar{A}$ domains, 
because further collapse of the $\bar{A}$ also yields unfavorable 
domain wall $W_{A-B}$.  
Therefore the collapse of three domains of $\bar{B}, \bar{A}$, and $B$ 
will occur concurrently. 
This is a very rare event especially at low temperatures, because the 
domain size becomes larger and larger as the temperature is decreased 
toward $T_{C2}$. The reverse is also true. That is, the concurrence 
of three domains in the $\langle 2 \rangle$ phase is also very rare event.
These properties explain a well-known phenomenon of the 2D ANNNI model, i.e., 
a huge number of the MC sweep is necessary to get an equilibrium spin 
configuration.

Also development of the spin structure exhibits an interesting property 
in this temperature range. 
Figure 10 shows the development of the maximum spin overlap $q_m(t)$ 
starting with paramagnetic spin configuration for various temperatures. 
Here we adopt a single-spin-flip heat-bath MC algorithm and $t$ being 
the number of the MC sweep.  
For a temperature well-above $T_{C2}$, $q_m(t)$ monotonously increases toward 
$q_m(\infty)$, which is estimated in the equilibrium CHB MC simulation. 
On the other hand, for temperatures $T \sim T_{C2}$, $q_m(t)$ exhibits 
two-stage development. In the first stage, it increases algebraically with $t$ 
and reach a value of $\bar{q_m} \sim 0.5$ at $\bar{t} \sim 10^4$, which are 
almost independent of the temperature (even for well-below $T_{C2}$).
In the second stage, $q_m(t)$ slowly increases with $t$ until $q_m(t)$ 
reaches its equilibrium value $q_m(\infty)$. Again we see $q_m(t)$ are 
almost independent of the temperature within this time scale. 
We find that this two-stage development of $q_m(t)$ comes from the domain 
structure of the model. Figure 11 shows the $t$ dependence of $S(k)$ at 
$T \sim T_{C2}$. As $t$ increases from $t = 0$, 
the peak of $S(k)$ develops and becomes of a single-peaked at $t \sim 10^4$. 
Above this time, the peak position $\bar{k}$ increases very slowly toward 
to $k_p$ of the equilibrium result with clarifying its shape. 
That is, the first stage of the development of the spin structure is the 
creation of some quasi-periodic spin arrangement, and in the second stage 
the period of the periodic structure is gradually changes to fit its 
equilibrium one. The later stage is the collapse of different domains 
which is very slow as discussed above.

\begin{figure}
\includegraphics{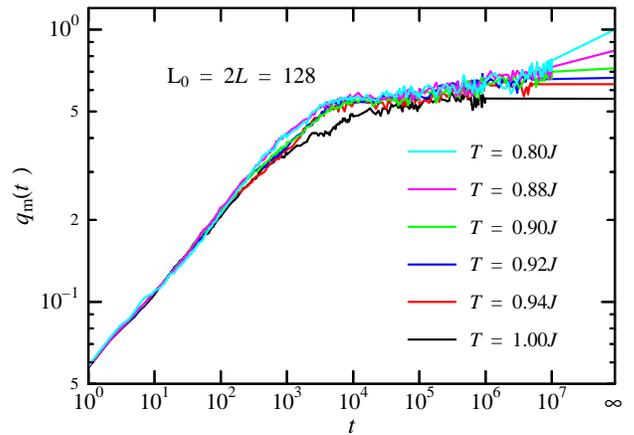}\\
\caption{(color online) 
The MC sweep $t$ dependence of the maximum spin overlap $q_m(t)$ starting 
with paramagnetic spin configuration in the ANNNI model with $\kappa = 0.6$. 
The last points are ones estimated by the equilibrium CHB simulation. 
}
\end{figure}

\begin{figure}
\includegraphics{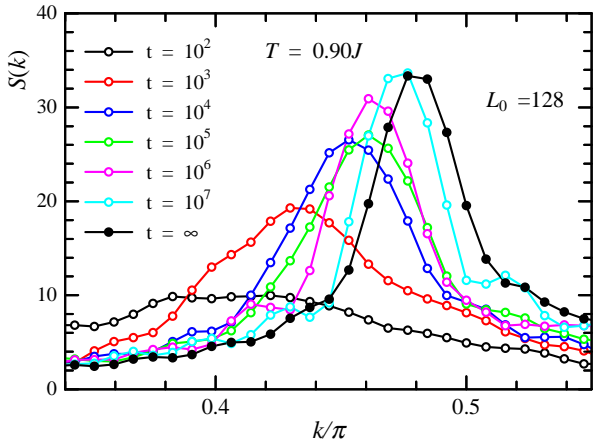}\\
\caption{(color online) 
The MC sweep $t$ dependence of the Fourier component $S(k)$ of the 
spin structure starting with paramagnetic spin configuration 
in the ANNNI model with $\kappa = 0.6$. 
$S(k)$ for $t = \infty$ is one estimated by the equilibrium CHB 
simulation. 
}
\end{figure}


\begin{acknowledgments}
We are thankful for the fruitful discussions with Professor S. Fujiki. 
Part of the results in this research was obtained
 using supercomputing resources
at Cyberscience Center, Tohoku University.
\end{acknowledgments}


%

\end{document}